# Superconductivity induced by doping Rh in $CaFe_{2-x}Rh_xAs_2$

Yanpeng Qi, Lei Wang, Zhaoshun Gao, Dongliang Wang, Xianping Zhang, Chunlei Wang, Chao Yao, Yanwei Ma*

Key Laboratory of Applied Superconductivity, Institute of Electrical Engineering, Chinese Academy of Sciences, P. O. Box 2703, Beijing 100190, China

**Abstract:**

In this paper we report the synthesis of iron-based superconductors $CaFe_{2-x}Rh_xAs_2$ using one-step solid state reaction method, which crystallizes in the $ThCr_2Si_2$–type structure with a space group I4/mmm. The systematic evolution of the lattice constants demonstrates that the Fe ions are successfully replaced by the Rh. By increasing the doping content of Rh, the spin-density-wave (SDW) transition in the parent compound is suppressed and superconductivity emerges. The maximum superconducting transition temperature is found at 18.5 K with the doping level of x = 0.15. The temperature dependence of DC magnetization confirms superconducting transitions at around 15 K. The general phase diagram was obtained and found to be similar to the case of Rh-doping Sr122 system. Our results explicitly demonstrate the feasibility of inducing superconductivity in Ca122 compounds by higher d-orbital electrons doping, however, different Rh-doping effect between FeAs122 compounds and FeAs1111 systems still remains an open question.

* Author to whom correspondence should be addressed; E-mail: ywma@mail.iee.ac.cn

## 1. Introduction

The discovery of iron pnictide superconductors [1] has broken cuprate "monopoly" in the physics of high-temperature superconductivity (HTSC) compounds and revived the hopes both on further progress in this field related to the synthesis of new perspective high temperature superconductors, as well as on more deep understanding of mechanisms of HTSC. Up to now, five homologous series of iron-based superconductors have been discovered, commonly denoted as 1111 phase (REFeAsO with RE = rare earth) [2-6], 122 phase ($AEFe_2As_2$ with AE = alkaline earth) [7-10], 111 phase (AFeAs with A = alkali metal) [11, 12] and 11 phase (FeTe or FeSe) [13], 42622 phase ($Sr_4Sc_2O_6Fe_2P_2$, $Sr_4V_2O_6Fe_2As_2$ et al.)[14,15]. In these compounds, the Fe-As layer is thought to be responsible for superconductivity, which is separated by different carrier reservoir layer such as RE-O, AE, A and so on.

The parent $CaFe_2As_2$ has the tetragonal ThCrSi-type crystal structure with a space group I4/mmm, similar to other member of the $AeFe_2As_2$ family. At ambient pressure $CaFe_2As_2$ undergoes a transition from a non-magnetically ordered tetragonal to an antiferromagnetic orthorhombic phase [16, 17]. Doping or applied pressure could suppress this transition and induce the superconductivity [18- 20]. In this paper, we investigate Rh-doping effect in Ca122 system. Rhodium is a 4d metal, which locates just below Co and above Ir in the periodic table of elements. For Rh-doping a suppression of the SDW state and emergence of superconductivity have been reported in (Ba, Sr)$Fe_2As_2$ and Re1111 system [21-25]. Now we report the successful synthesis of the new superconductor $CaFe_{2-x}Rh_xAs_2$ with a $T_c$ of 18 K, X-ray diffraction indicates that the material has formed the $ThCr_2Si_2$–type structure with a space group I4/mmm. Resistivity, DC magnetic susceptibility as well as the phase diagram has been determined in the system of $CaFe_{2-x}Rh_xAs_2$.

## 2. Experimental

We employed a one-step solid state reaction method to synthesize the $CaFe_{2-x}Rh_xAs_2$ samples. The details of fabrication process are described elsewhere [6, 10]. Stoichiometric amounts of the starting elements Ca, Fe, Rh and As were thoroughly grounded and encased into pure Nb tubes. After packing, this tube was subsequently rotary swaged and sealed in a Fe tube. The sealed samples were heated to 850 $^{o}$C and kept at this temperature for 35 hours. Then it was cooled down slowly to room temperature. The high purity argon gas was allowed to flow into the furnace during the heat-treatment process. The sintered samples were obtained by breaking the Nb tube. It is noted that all the weighing, mixing and encasing procedures were performed in a glove box in which high pure argon atmosphere is filled.

The X-ray diffraction measurement was performed at room temperature using an MXP18A-HF-type diffractometer with Cu- $K_{\alpha}$ radiation from 20° to 80° with a step of 0.01°. The analysis of x-ray powder diffraction data was done and the lattice constants were derived. DC magnetization measurement was carried out on a Quantum Design physical property measurement system (VSM, PPMS). The resistance data were collected on the Quantum Design instrument physical property measurement system (Quantum Design, PPMS-9T). Rectangular specimens with dimensions of about 8 × 3 × 2 mm$^3$ were cut from the samples and resistivity measurements were performed by the conventional four-point-probe method. The electric contacts were made using silver paste with the contacting resistance below 0.05 at room temperature. The data acquisition was done using a DC mode of the PPMS, which measures the voltage under an alternative DC current and the sample resistivity is obtained by averaging these signals at each temperature.

**3. Results and discussion**

In order to have a comprehensive understanding of the evolution induced by the doping process, we have measured the X-ray diffraction patterns for all samples. Figure 1 shows the representative XRD patterns of $CaFe_{2-x}Rh_xAs_2$ samples (x=0.0 and 0.1). The XRD peaks can be well indexed based on a tetragonal cell of $ThCr_2Si_2$–type structure, indicating that the samples are essentially single phase. Small amount of impurity phases, mostly perhaps FeAs, were also observed in the XRD patterns. By fitting the data to the structure calculated with the software X'Pert Plus, we get the lattice constants. The lattice parameters are plotted in Fig. 2 as a function of Rh content. In the parent phase $CaFe_2As_2$, the lattice constants are a = 0.3887 nm and c = 1.1731 nm, which are in agreement with previous reports [9, 16-17]. With increasing

Rh content, the *a*-axis lattice constant expands a bit, while *c*-axis shrinks significantly. The change of lattice parameters of in $CaFe_{2-x}Rh_xAs_2$ compounds is similar to the case of doping the Fe with Ru, Ir or Pd in iron based superconductors [21-22, 26-28]. The nearly linear variation in the lattice parameters indicates a successful chemical substitution.

Figure 3 shows the temperature dependence of the electrical resistivity for $CaFe_{2-x}Rh_xAs_2$ samples with x = 0 to 0.3 respectively. Data are shown normalized by the room temperature resistivity $R_{300K}$, to remove uncertainty in estimates of the absolute value due to geometric factors. Successive data sets are offset vertically by 1 for clarity. The parent phase exhibits a resistivity anomaly at about 165 K, which associated with the magnetic/structural phase transition. The anomaly is clearly suppressed in temperature with increasing x. For the samples with x > 0.05 superconductivity is evident from a drop in resistivity, while the anomaly associated with structural/magnetic phase transition is still observable in the normal state. With a further increase in Rh fraction, the resistivity anomaly disappeared completely and the superconducting critical temperature Tc reaches the highest value of 18.5 K in the sample with x = 0.15, which is comparable with the value of Rh-doping (Ba,Sr)122 system[21, 22]. The figure 5a shows an enlarged plot of R versus T for $CaFe_{1.85}Ru_{0.15}As_2$ at low temperature. It is clear that the resistivity vanishes at about 7 K. It is noted that the resistivity drop was not converted to an uprising with the Rh doping, which is different from the Co or Pd doped samples [21, 29]. This difference may be induced by the two effects which give opposite contributions to the resistivity in the system: the decrease of the scattering rate as well as the charge carrier densities. In addition, the normal state resistivity of the superconducting sample shows a roughly linear behavior starting just above Tc all the way up to 300K.

Figure 4 shows the temperature dependence of the DC magnetization for the $CaFe_{2-x}Rh_xAs_2$ samples. The measurement was carried out under a magnetic field of 20 Oe in the zero-field-cooled and field-cooled processes. We compare the diamagnetic signals in the samples with x = 0.05, 0.10, 0.15, 0.20, 0.25. A clear diamagnetic signal appears around 15 K in the samples, which correspond to the middle transition temperature of the resistivity data. The figure 5b shows M-T pattern for the sample x = 0.15. The diamagnetism and presence of zero resistance in the samples are the proof that Rh substitution in the $CaFe_2As_2$ compounds lead to superconductivity. Although the connectivity between grains in our polycrystalline samples as well as the vortex pinning effect give same influence on the diamagnetization signal, the strong diamagnetization value certainly indicate a rather large volume of superconductivity.

It is well known that $CaFe_2As_2$ undergoes a transition from a non-magnetically ordered tetragonal to an antiferromagnetic orthorhombic phase at ambient pressure, however, neutron powder diffraction measurements of $CaFe_2As_2$ under hydrostatic pressure found that for p > 0.35 GPa (at T = 50 K), the antiferromagnetic orthorhombic phase transforms to a new, non-magnetically ordered, collapsed tetragonal structure. When the sample is cooled across the tetragonal-collapsed tetragonal phase transition there is an extremely anisotropic change in the unit cell dimensions: the a-axis expands by 2.5% and the c-axis contracts by 9% [9, 16-20]. However, lattice parameters variation are found in Rh doping Ca122 system, take the compound x = 0.3 for example, *c*-axis lattice parameter decrease is 2 %, however, *a*-axis lattice parameters increase only about 1%, we suspect that the *c*-axis shrinkage effect occurred in this Ca122 system could add charge carriers as well as chemical pressure, however, which of these effects is more important for superconductivity requires further exploration as does their possible interplay.

Based on the measurements described above, we can establish a composition-temperature phase diagram for $CaFe_{2-x}Rh_xAs_2$, shown in Fig. 6. Both $T_{an}$ and $T_c$ were defined as the temperature which the anomaly appears in resistivity and the superconducting transition, respectively. With increasing Rh-doping, the temperature of the anomaly is driven down, and the superconducting state emerges at x = 0.05, and reaching a highest Tc of 18.5 K at x = 0.15, the superconducting state continues from x = 0.05, and even appears at the doping level of x = 0.3. This general phase diagram looks very similar to that of Co doping [29]. Considering Rh locates just below Co and above Ir in the periodic table of elements, we would conclude that the superconductivity induced by Rh doping shares the similarity as that of Co or Ir doping. It should be noted that there exists a region in which the SDW and superconductivity coexist. This result is different from Rh-doping FeAs1111 superconductors, in which no coexistence of antiferromagnetisim and superconductivity in an underdoped region is found [24]. Thus further experimental and theoretical work is needed for the interpretation of the different properties between the FeAs1111 type superconductors and the FeAs122 type superconductors.

## 4. Conclusions

In summary, we have synthesized a series of layered $CaFe_{2-x}Rh_xAs_2$ compounds with the $ThCr_2Si_2$–type structure. The systematic evolution of the lattice constants indicates successful Rh-doping into the lattice. The diamagnetism and presence of zero resistance in the samples are the proof that Rh substitution in the $CaFe_2As_2$ compounds lead to superconductivity. A detailed phase diagram with evolution from SDW to superconducting state with Rh doping is given.

## Acknowledgments

The authors thank Profs. Haihu Wen, Liye Xiao and Liangzhen Lin for their help and useful discussions. This work is partially supported by the Beijing Municipal Science and Technology Commission under Grant No. Z09010300820907, National '973' Program (Grant No. 2011CBA00105) and the National Natural Science Foundation of China (Grant No. 51025726 and 51002150).

## References

[1] Y. Kamihara, T. Watanabe, M. Hirano and H. Hosono, *J. Am. Chem. Soc.* **130**, 3296 (2008).

[2] X. H. Chen, T. Wu, G. Wu, R. H. Liu, H. Chen and D. F. Fang, *Nature* **453**, 376 (2008).

[3] G. F. Chen, Z. Li, D. Wu, G. Li, W. Z. Hu, J. Dong, P. Zheng, J. L. Luo and N. L. Wang, *Phys. Rev. Lett.* **100**, 247002 (2008).

[4] H. H. Wen, G. Mu, L. Fang, H. Yang and X. Zhu, *Europhys. Lett.* **82**, 17009 (2008).

[5] Z. A. Ren, J. Yang, W. Lu, W. Yi, X. L. Shen, Z. C. Li, G. C. Che, X. L. Dong, L. L. Sun, F. Zhou and Z. X. Zhao, *Chin. Phys. Lett.* **25**, 2215 (2008).

[6] Y. P. Qi, Z. S. Gao, L. Wang, D. L. Wang, X. P. Zhang and Y. W. Ma, *Supercond. Sci. Technol.* **21**, 115016 (2008).

[7] M. Rotter, M. Tegel and D. Johrendt, *Phys. Rev. Lett.* **101**, 107006 (2008).

[8] K. Sasmal, B. Lv, B. Lorenz, A. Guloy, F. Chen, Y. Xue and C. W. Chu, *Phys. Rev. Lett.* **101**, 107007 (2008).

[9] N. Ni, S. Nandi, A. Kracher, A. I. Goldman, E. D. Mun, S. L. Bud'ko and P. C. Canfield, *Phys. Rev. B* **78**, 014523 (2008).

[10] Y. P. Qi, Z. S. Gao, L. Wang, D. L. Wang, X. P. Zhang and Y. W. Ma, *New J. Phys.* **10,** 123003 (2008).

[11] X. C. Wang, Q. Q. Liu, Y. X. Lv, W. B. Gao, L. X. Yang, R. C. Yu, F. Y. Li, C. Q. Jin, *Solid State Commun.* **148**, 538 (2008).

[12] J. H. Tapp, Z. J. Tang, B. Lv, K. Sasmal, B. Lorenz, Paul C. W. Chu and A. M. Guloy, *Phys. Rev. B* **78**, 060505(R) (2008).

[13] F. C. Hsu, J. Y. Luo, K. W. Yeh, T. K. Chen, T. W. Huang, P. M. Wu, Y. C. Lee, Y. L. Huang, Y. Y. Chu, D. C. Yan, M. K. Wu, *PNAS* **105**, 14262 (2008).

[14] H. Ogino, Y. Matsumura, Y. Katsura, K. Ushiyama, S. Horii, K. Kishio and J Shimoyama, *Supercond. Sci. Technol.* **22**, 075008 (2009).

[15] X. Zhu, F. Han, G. Mu, P. Cheng, B. Shen, B. Zeng, and H-H Wen, *Phys. Rev. B*

**79**, 220512(R) (2009).

[16] G. Wu, H. Chen, T. Wu, Y. L. Xie, Y. J. Yan, R. H. Liu, X. F.Wang, J. J. Ying and X. H. Chen, *J. Phys.: Condens. Matter* **20**, 422201 (2008).

[17] F. Ronning, T. K. limczuk, E. D. Bauer, H. Volz and J D Thompson, *J. Phys.: Condens. Matter*, **20**, 322201 (2008).

[18] M. S. Torikachvili, S. L. Bud'ko, N. Ni, P. C. Canfield, *Phys. Rev. Lett.* **101**, 057006 (2008).

[19] N. Kumar, R. Nagalakshmi, R. Kulkarni, P. L. Paulose, A. K. Nigam, S. K. Dhar and A. Thamizhavel, *Phys. Rev. B* **79**, 012504 (2009).

[20] P.C. Canfield, S.L. Bud'ko, N. Ni, A. Kreyssig, A.I. Goldman, R.J. McQueeney, M.S. Torikachvili, D.N. Argyriou, G. Luke, W. Yu，*Physica* C **469,** 404 (2009).

[21] F. Han, X. Zhu, P. Cheng, G. Mu, Y. Jia, L. Fang, Y. L. Wang, H. Luo, B. Zeng, B. Shen, L. Shan, C. Ren and H. H. Wen, *Phys. Rev. B* **80**, 024506 (2009).

[22] N. Ni, A. Thaler, A. Kracher, J. Q. Yan, S. L. Bud'ko and P. C. Canfield, *Phys. Rev. B* **80**, 024511 (2009).

[23] Y. P. Qi, L. Wang, Z. S. Gao, D. L. Wang, X. P. Zhang Z. Y. Zhang and Y. W. Ma, *Europhys. Lett.* **89**, 67007 (2010).

[24] H. Y. Shi, X. L. Wang, T. L. Xia, Q. M. Zhang, X. Q. Wang, T.S. Zhao，*Phys. Status Solidi* RRL **4**, 67 (2010).

[25] S. Muir, A. W. Sleight, M. A. Subramanian, *Materials Research Bulletin* **45,** 392 (2010).

[26] Y. P. Qi, L. Wang, Z. S. Gao, D. L. Wang, X. P. Zhang and Y. W. Ma, *Physica* C **469,** 1921 (2009).

[27] S. Paulraj, S. Sharma, A. Bharathi, A. T. Satya, S. Chandra, Y. Hariharan and C. S. Sundar, Cond-mat: arXiv, 0902.2728 (2009).

[28] Y. P. Qi, L. Wang, Z. S. Gao, D. L. Wang, X. P. Zhang and Y. W. Ma, *Phys. Rev. B* **80**, 054502 (2009).

[29] J. H. Chu, J. G. Analytis, C. Kucharczyk, I. R. Fisher, *Phys. Rev. B* **79**, 014506 (2009).

[30] Ma Y W, Gao Z S, Wang L, Qi Y P, et al., *Chin. Phys. Lett.* **26** 037401 2009

## Captions

Figure 1 (Color online) XRD patterns of the $CaFe_{2-x}Rh_xAs_2$ samples. The impurity phases are marked by *.

Figure 2 (Color online) Lattice parameters as function of Rh content. It is clear that the *a*-axis lattice expands, while *c*-axis lattice shrinks with Rh substitution.

Figure 3 (Color online) Temperature dependence of resistivity for the $CaFe_{2-x}Rh_xAs_2$ samples measured in zero field. The data are normalized to $R_{300K.}$

Figure 4 (Color online) Temperature dependence of DC magnetization for the $CaFe_{2-x}Rh_xAs_2$ samples. The measurements were done under a magnetic field of 20 Oe with zero-field-cooled and field-cooled modes.

Figure 5 (Color online) Temperature dependence of resistivity (figure 5a) and DC magnetization (figure 5b) at low temperature for the $CaFe_{0.85}Rh_{0.15}As_2$ samples.

Figure 6 (Color online) Electronic phase diagram for $CaFe_{2-x}Rh_xAs_2$. $T_{an}$ and $T_c$ denote the resistivity anomaly temperature and the critical transition temperature, respectively. The dashed line provides a guide to the eyes for the possible SDW / structural transitions near the optimal doping level.

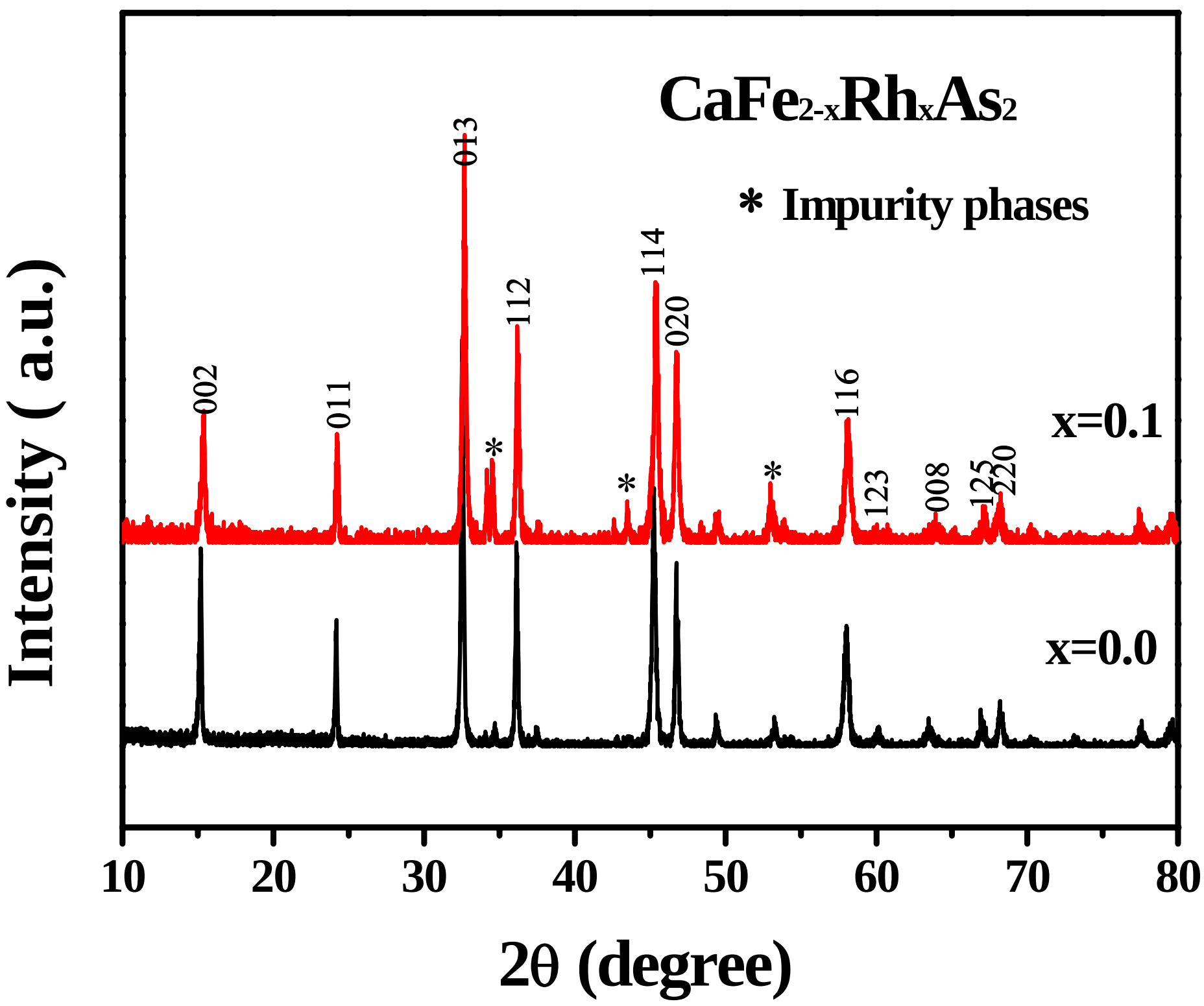

Fig.1 Qi et al.

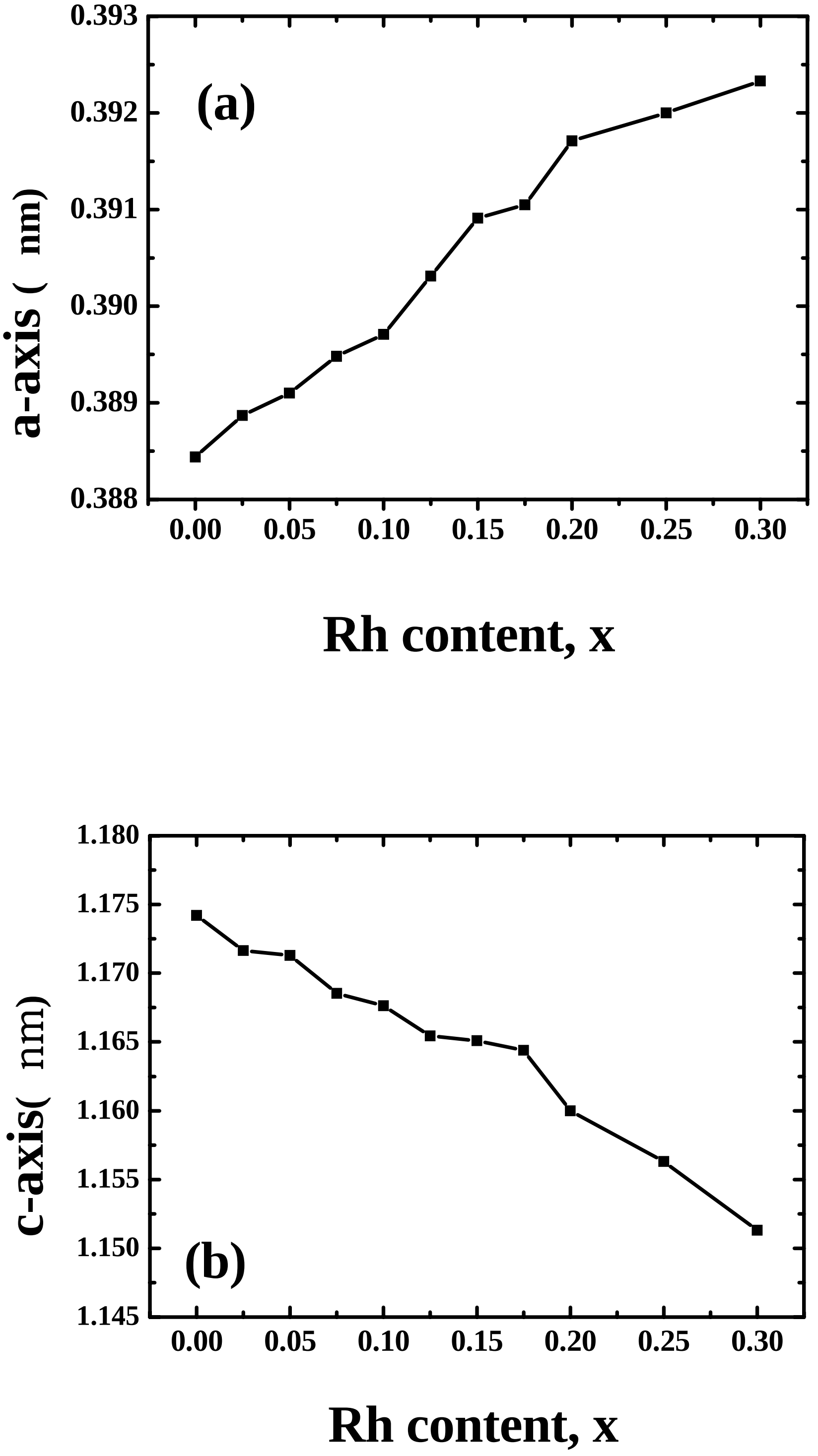

Fig.2 Qi et al.

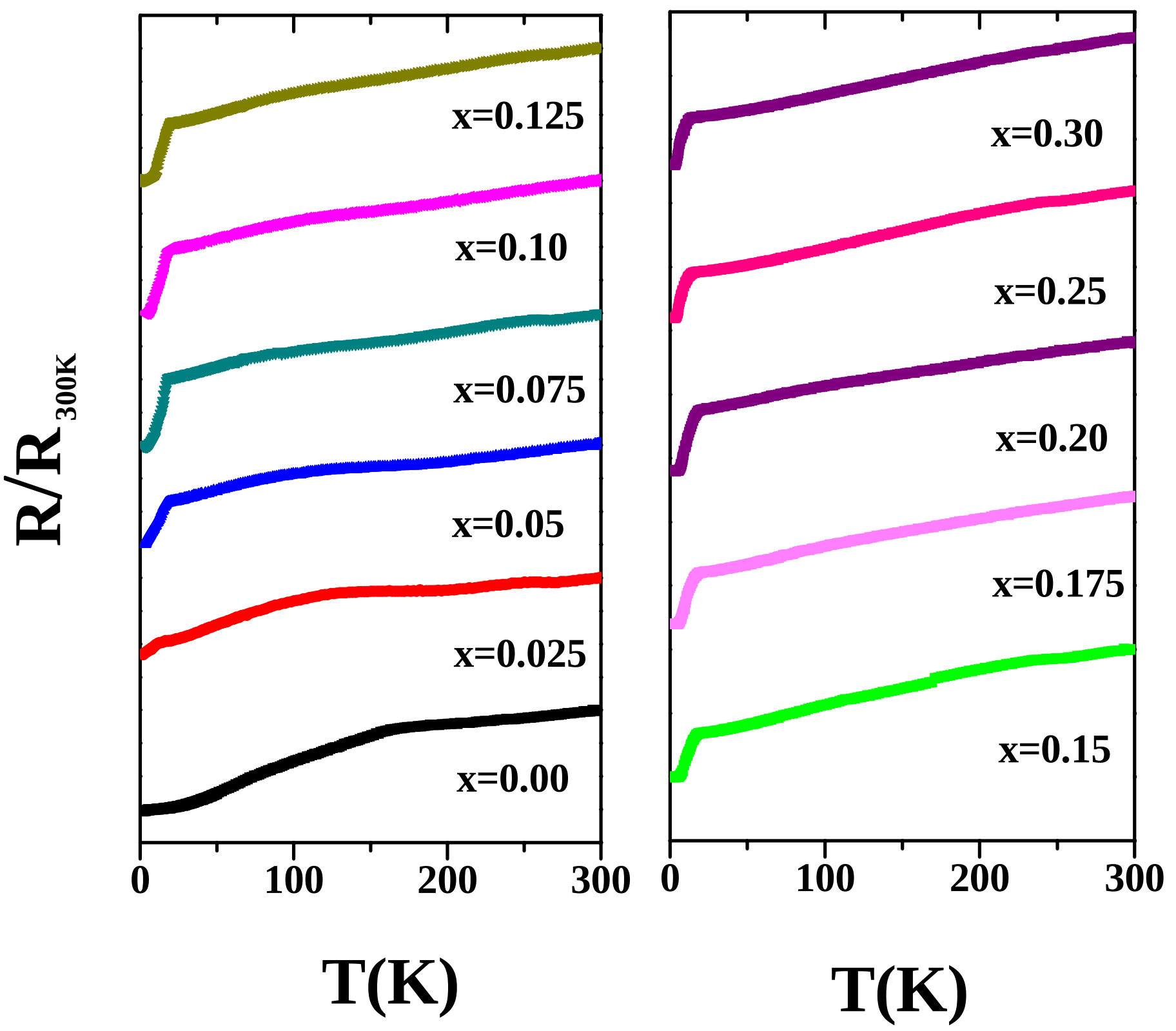

Fig.3 Qi et al.

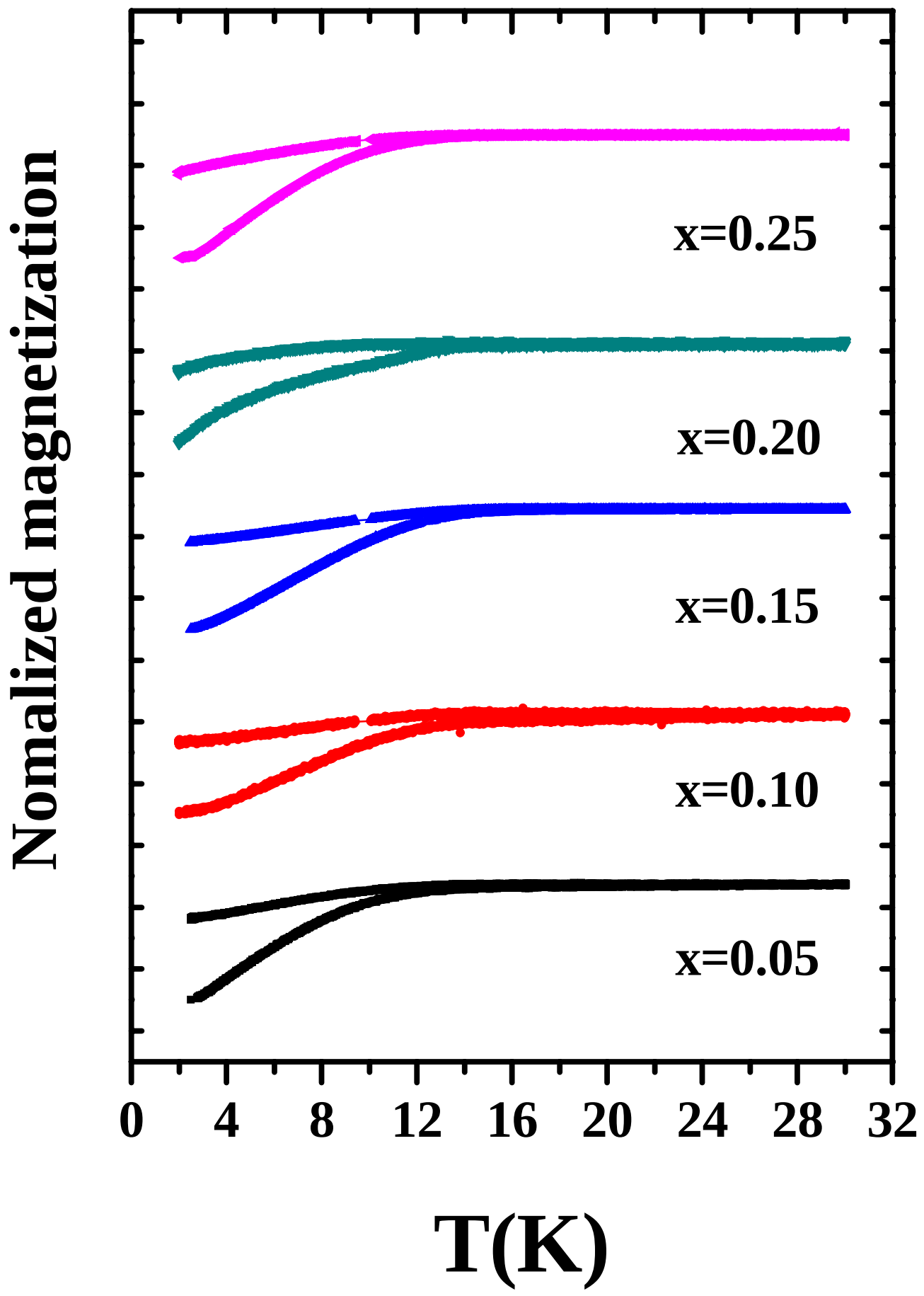

Fig.4 Qi et al.

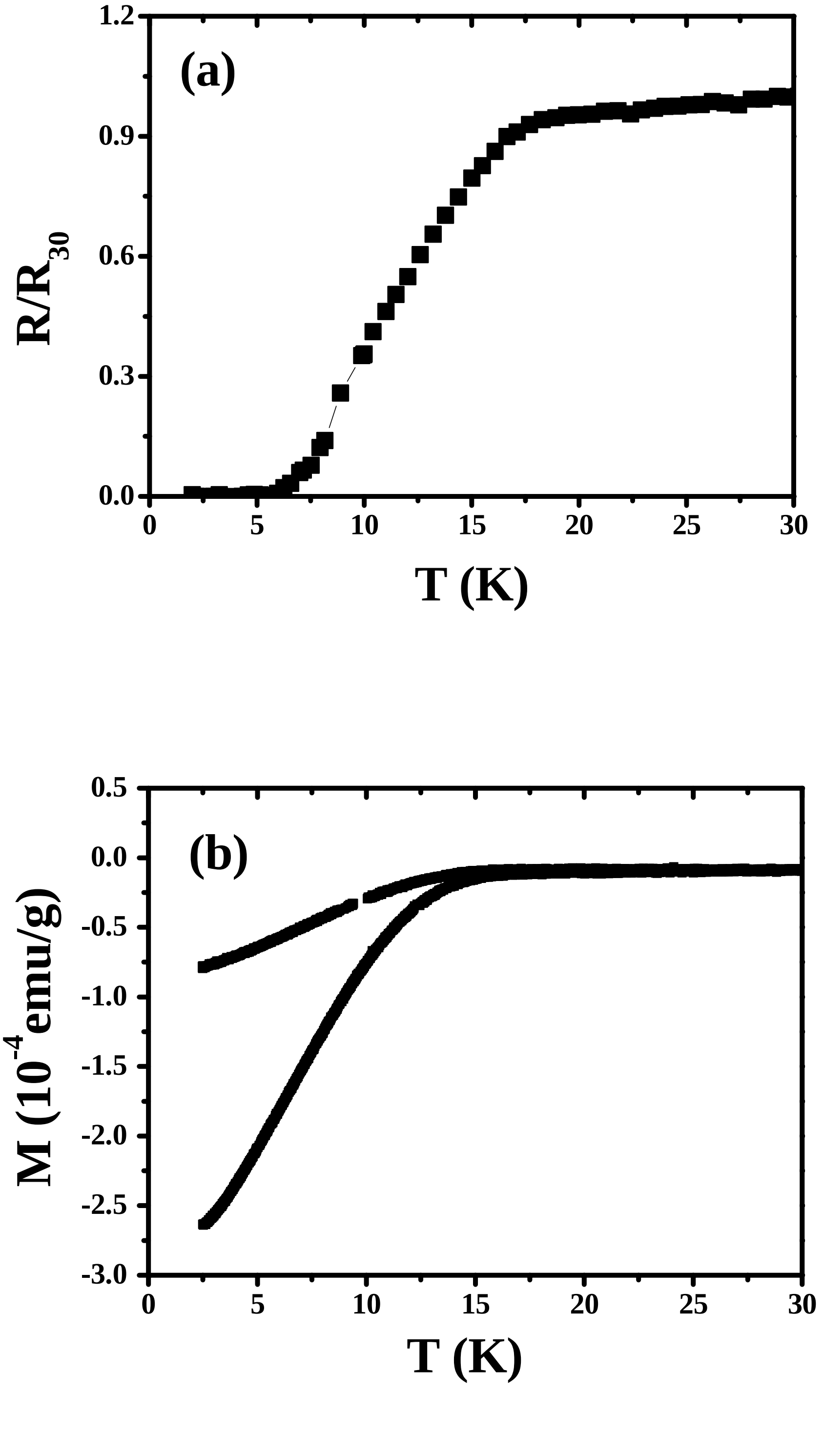

Fig.5 Qi et al.

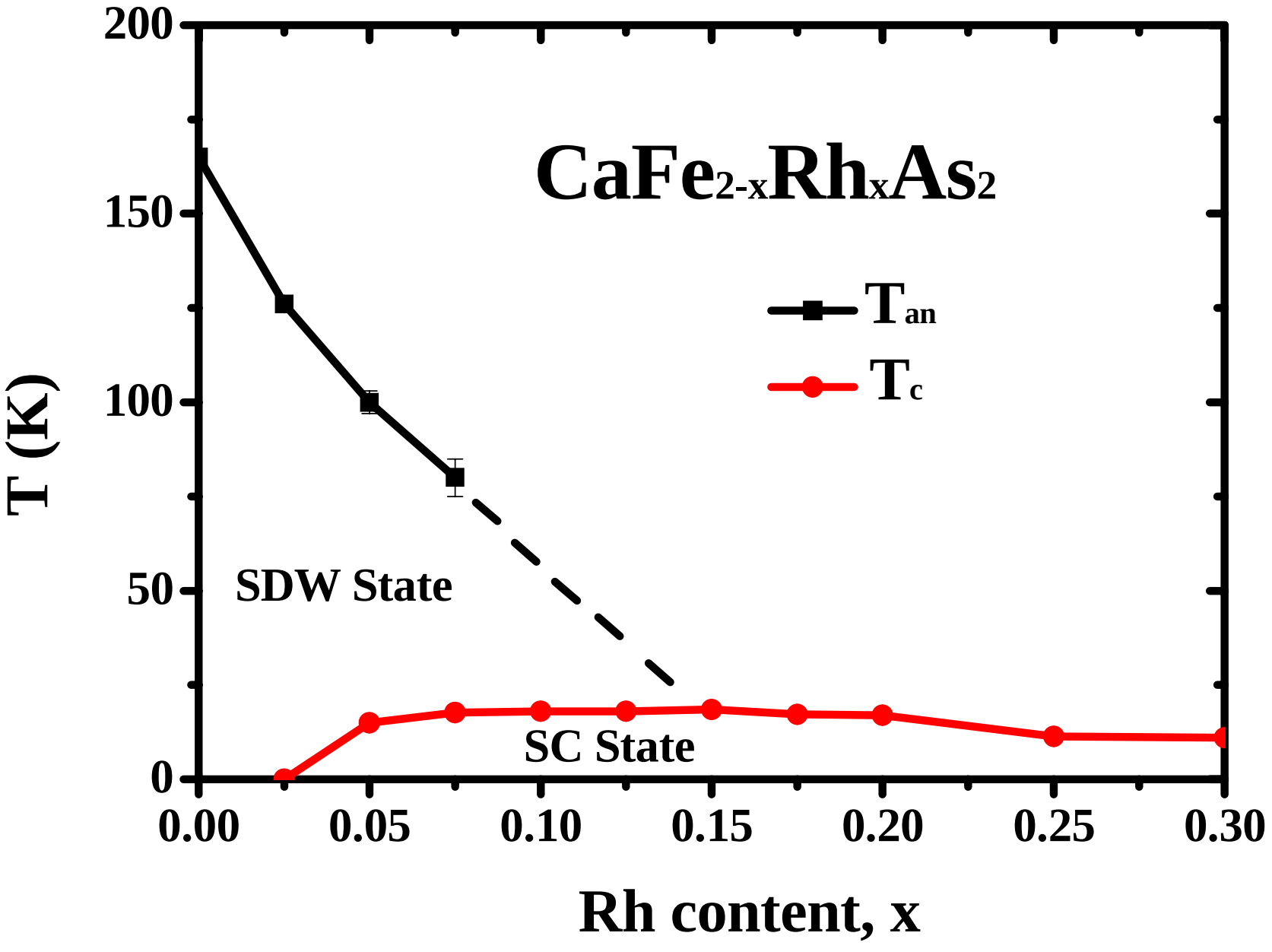

Fig.6 Qi et al.